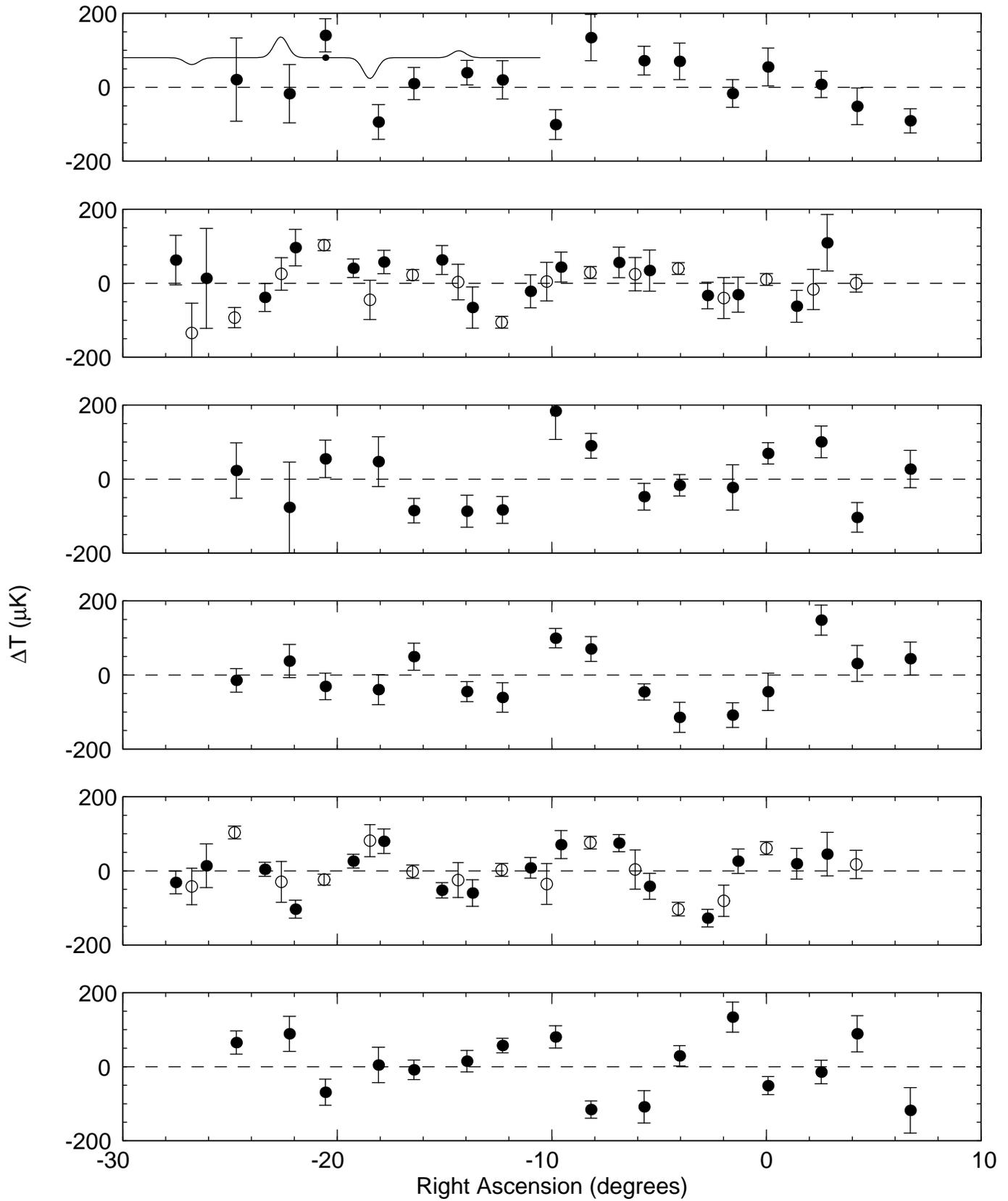

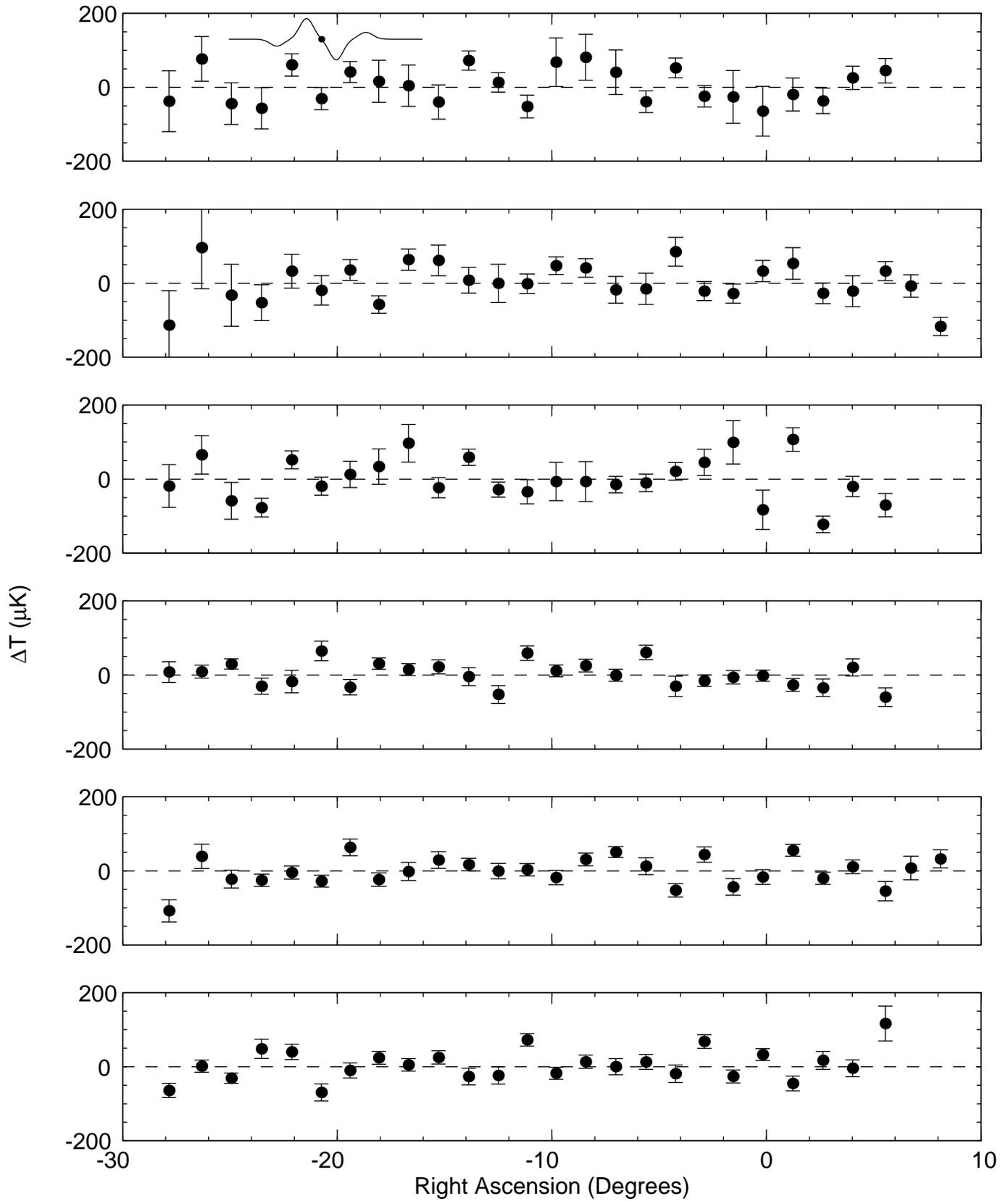

# ANISOTROPY IN THE MICROWAVE SKY AT 90 GHZ: RESULTS FROM PYTHON III

S. R. Platt[1,2], J. Kovac[3], M. Dragovan[2,3], J. B. Peterson[4], and J. E. Ruhl[5]

## ABSTRACT

The third year of observations with the Python microwave background experiment densely sample a 5.5° × 22° region of sky that includes the fields measured during the first two years of observations with this instrument. The sky is sampled in two multipole bands centered at $\ell \approx 92$ and $\ell \approx 177$. These two data sets are analyzed to place limits on fluctuations in the microwave sky at 90 GHz. Interpreting the observed fluctuations as anisotropy in the cosmic microwave background we find flat band power estimates of $\delta T_\ell \equiv \sqrt{\ell(\ell+1)C_\ell/(2\pi)} = 54^{+14}_{-12}$ μK at $\ell = 92^{+37}_{-24}$ and $\delta T_\ell = 58^{+15}_{-13}$ μK at $\ell = 177^{+66}_{-58}$. Combining the entire three year set of Python observations, we find that the angular power spectrum of fluctuations has a spectral index m = $0.02^{+.18}_{-.16}$ and an amplitude $\delta T_{\ell_e} = 53^{+13}_{-11}$ μK at $\ell_e = 145^{+59}_{-74}$ for the functional form $\delta T_\ell = \delta T_{\ell_e}(\ell/\ell_e)^m$. The stated uncertainties in the amplitudes and spectral index represent 1σ confidence intervals in the likelihood added in quadrature with a 20% calibration uncertainty and an estimate of the effects introduced due to imperfect overlap of the beams on the sky. The limits of $\ell$ are determined from the full width at half maximum of the window functions.

Subject headings: cosmic microwave background - cosmology: observations

[1]Yerkes Observatory, University of Chicago, Williams Bay, WI 53191-0258

[2]Dept. of Astronomy and Astrophysics, University of Chicago, Chicago, IL 60637

[3]Enrico Fermi Institute, University of Chicago, Chicago, IL 60637

[4]Dept. of Physics and Astronomy, Carnegie-Mellon University, Pittsburgh, PA 15213

[5]Dept. of Physics, University of California, Santa Barbara, Santa Barbara, CA 93106

## 1. INTRODUCTION

Primordial density fluctuations in the early universe have evolved to form the large-scale structure observed today. Many models have been developed that attempt to explain the details of this evolution (White, Scott & Silk 1994). Anisotropy measurements of the cosmic microwave background (CMB) at a range of angular scales are one of the few direct probes of the conditions and processes in the early universe that can be used to constrain these models. Measurements at large angular scales (Bennett et al. 1994; Ganga et al. 1993) have been used to normalize the cosmological power spectrum. The amplitude and shape of the spectrum at smaller angular scales, however, is still not well-determined although a number of groups (e.g. Cheng et al. 1994; Dragovan et al. 1994; Netterfield et al. 1995; Gundersen et al. 1995; Tanaka et al. 1996) have reported detections of anisotropy on angular scales near one degree.

The first observations (hereafter PyI) with the Python microwave background anisotropy experiment yielded a statistically significant detection of anisotropy (Dragovan et al. 1994) at an angular scale of $\ell = 92$. Sixteen sky positions were observed. These results were confirmed with observations of the same sky positions the following year (hereafter PyII). At that time, fifteen additional sky positions were also measured (Ruhl et al. 1995a).

To increase sky and angular scale coverage we returned to the Amundsen-Scott South Pole station for the third year of observations (hereafter PyIII) with this instrument (Platt et al. 1995). We report here the results of these observations, made between 1994 November 15 and 1995 January 1. The observations were made of a $5.5° \times 22°$ region of sky that includes the PyI and PyII fields. Measurements of several of the individual PyI and PyII sky positions reconfirmed our past results and provided a check for our pointing. Using our original observing strategy, 96 new sky positions were observed. This same region of sky was also observed at 154 positions with a smaller chopper differencing angle to measure fluctuations at an angular scale of $\ell = 177$.

## 2. INSTRUMENT

All of the Python measurements have been made at a frequency of 90 GHz[†] with a five detector bolometric photometer mounted on a 0.75 m diameter off-axis parabolic telescope. Four of the bolometers are sensitive to signals from the sky, while one is placed in a dark cavity and is used as a monitor of extraneous pickup. The four optical channels are scanned horizontally across the sky in a three point pattern by rotation of a vertical flat mirror at 2.5 Hz. This switching pattern ensures that the measurements are made at a constant elevation, minimizing changes in offsets due to the atmosphere. Two ground screens, one rotating with the telescope and the other fixed to the ground, surround the instrument and telescope to block radiation from the ground and Sun.

Each detector consists of a neutron transmutation doped germanium chip glued to diamond wafer substrate with an absorbing layer of bismuth evaporated onto it. The four optical detectors are arranged in a $2 \times 2$ square array and are coupled to the sky through identical sets of single-mode waveguide filters and corrugated feed horns. When the photometer is mounted on the telescope, the detector array is aligned such that each pair of detectors in a given row observes the same elevation, with the two rows separated by 2.75°. The horizontal separation between detectors is also 2.75° on the sky. The beam profiles are well-approximated by a Gaussian with a full-width at half maximum (FWHM) of $0.75° \pm 0.05°$. The bolometers are cooled to 50 mK with a $^3$He guarded Adiabatic Demagnetization Refrigerator (Ruhl & Dragovan 1992).

During the PyII observing season, one of the four optical channels did not work. At the end of that season, this detector was replaced and an electrical problem inside the dewar repaired. All four optical detectors were operational during the PyIII season, although one detector remained a factor of three less sensitive than the others. Unfortunately, a new problem developed with the dark channel after these repairs were completed. Excess electrical noise in this channel required the use of a gain that was ten times lower than for the optical channels, limiting the usefulness of this detector as a diagnostic tool. There is no evidence that the noise in the dark channel contaminated the optical channels.

---

[†] During the 1995-96 Austral summer season, observations were made with a 43 GHz system on the Python telescope. The results will be presented in a forthcoming paper.

The entire Python experiment was relocated at the end of the PyII season from a remote site 2 km from South Pole station to new, permanent facilities 1 km from the Station. The telescope and shields were reinstalled atop a 12 foot tower, and all of the alignment and pointing procedures were repeated.

## 3. OBSERVATIONS

Two series of observations of a 5.5° × 22° region of sky centered at $\alpha = 23.37^h$, $\delta = -49.44°$ (J2000.0) were made during the PyIII season. This entire region is densely sampled on a grid with a spacing of 0.92° on the sky in elevation and right ascension. Because the beams are 0.75° FWHM, the sky is not fully sampled. We use instead the term densely sampled as fully sampled generally refers to FWHM sample spacing. The absolute telescope pointing error is estimated to be 0.1° as determined by measurements of the Moon and the Carinae nebula at $\alpha = 10.73^h$, $\delta = -59.65°$ (J2000.0).

The basic observing strategy used for all of the Python measurements is described in detail in Ruhl et al.(1995b). Fast 3-beam chopping (2.5 Hz full cycle) of the vertical flat mirror is combined with slow azimuthal beamswitching (typically 0.10 to 0.025 Hz) of the entire telescope to produce a four-beam response to a sky signal,

$$\Delta T_j = -\frac{1}{4}T_j + \frac{3}{4}T_{j+1} - \frac{3}{4}T_{j+2} + \frac{1}{4}T_{j+3} \tag{1}$$

where $T_j$ is the antenna temperature of position j. The output of each detector is sampled at 100 Hz and demodulated in software.

The first set of observations this season (hereafter $PyIII_L$) used the same procedure as PyI and PyII. For these measurements the chopper throw and the azimuthal telescope beamswitching were both 2.75° on the sky. Six new fields were observed in this configuration. Each field is composed of 16 sky positions that are measured with a series of scans. Every scan consists of measuring $\Delta T_j$ three times successively at each of seven telescope beamswitch positions separated horizontally by 2.75° on the sky. Because the spacing between detectors in the array is also 2.75°, the first six sky positions measured by a given detector on the right side of the array are each in turn measured by the corresponding detector on the left side of the array. The first and last sky positions in each row are only measured by a single detector.

The PyI and PyII fields were not re-examined except for a few partial scans used to verify our past results and to provide an additional check on telescope pointing. The new fields were selected by offsetting the telescope in elevation and right ascension from the original PyI field in steps of one-third of the separation between the detectors in the array (0.92° on the sky). Two new fields were observed at each of three elevation offsets (-0.92°, 0.0°, and +0.92°). The fields at the same elevation as the PyI and PyII fields (elevation offset 0.0°) are offset in right ascension by -0.92° and -1.84° on the sky. The two fields at each new elevation are offset in right ascension by 0.0° and +1.84° on the sky.

The second set of observations this season (hereafter PyIII$_S$) was made with both the chopper throw and telescope beamswitch reduced to 0.92°. These scans consist of measuring the four-beam response $\Delta T_j$ three times successively at each of 12 telescope beamswitch positions. No sky position was measured by more than one detector within any individual scan. Interleaving the results for the left- and right-hand detectors yields 24 measured sky positions separated by 0.92° on the sky for each of the two rows of detectors in the array. Scans were made at three telescope elevation offsets (-0.92°, 0.0°, and +0.92°) chosen to overlap the PyI and PyIII$_L$ fields. At each elevation, scans were made offset in right ascension by 0.0° and -2.75° on the sky so that each sky position was measured at different times by a right-hand and left-hand detector. A total of 154 sky positions were measured in this way.

## 4. CALIBRATIONS

The DC calibration of the detectors was typically performed several times per dewar cycle using a combination of sky dips, low emissivity ambient temperature foam loads, and high emissivity liquid nitrogen and liquid oxygen cold loads (Dragovan et al. 1994; Ruhl et al. 1995a, 1995b). The DC response of the system was converted to an AC response by comparing the detector output produced while chopping slowly (0.1 Hz) on and off a foam load with that produced by chopping at the observing frequency (5 Hz). The relative gains of the detectors were checked approximately every twelve hours using the Carinae nebula.

The statistical uncertainty in the determination of the gains is of order 5% for each detector. The overall uncertainty in the gain, including systematic effects, is estimated to be 20%.

## 5. DATA ANALYSIS

A total of 100 hours of PyIII$_L$ and 125 hours of PyIII$_S$ data were collected in complete scans. Two statistics were used to remove scans contaminated by excessive noise on a channel by channel basis. Scans were grouped together according to the elevation and right ascension offsets used for the measurements, and each group was processed separately.

For each scan we calculate a mean and variance for the 3-beam sky signal for every telescope beamswitch position. The average 3-beam differences from neighboring telescope positions are then subtracted to yield the 4-beam measurements of $\Delta T_j$. The average of the variances, $\sigma_s^2$, is a measure of the noise on timescales corresponding to the integration time between telescope beamswitches, typically 10 - 40 seconds, and is dominated by instrument noise. A mean and variance are also calculated from the three 4-beam values of $\Delta T_j$ at each sky position for every detector. Because a new mean is calculated at every position, the average of these individual variances, $\sigma_m^2$, contains no contribution from celestial sources. This statistic is a good measure of noise on the timescale required to make the three 4-beam measurements of $\Delta T_j$, typically about 3 minutes.

Two passes were made through all complete scans within each group using the procedures described in Ruhl et al. (1995a) to remove scans with excessively high values of $\sigma_s$ or $\sigma_m$. The first pass removes scans with $\sigma_s$ greater than 1.5 times the peak of its distribution. The second pass removes scans with high values of $\sigma_m$. The level above which scans are cut is chosen to minimize the average final error bar in the binned data. Between 18% and 25% of the scans in PyIII$_L$ were cut in this way, depending on the detector. For PyIII$_S$, 8% - 17% of the scans were cut. The fraction of data cut was smaller for PyIII$_S$ than for PyIII$_L$ primarily because the smaller beamswitch and chopper throw reduced the sensitivity to larger angular scale atmospheric fluctuations.

The average values for $\sigma_s$ and $\sigma_m$ for all uncut scans in the two data sets are listed in Table 1. Both PyIII$_L$ and PyIII$_S$ show similar values for $\sigma_s$ which would be expected for a statistic dominated by instrument noise. For PyIII$_L$, $\sigma_m$ shows that the atmosphere is a significant source of noise, while the PyIII$_S$ scans are generally instrument noise limited. Table 1 also shows that the 3-beam offsets for PyIII$_S$ are significantly smaller than for PyIII$_L$.

Within every PyIII$_L$ scan, the left- and right-hand detectors in each row measure six common positions on the sky. A weighted average of the two measurements is used to combine these numbers into a single value at each sky position, as described in Ruhl et al. (1995a). An average value for every sky position is then determined by combining the overlapping measurements within each group of scans, and a statistical uncertainty assigned. These results are shown in Figure 1. The 96 new measurements are plotted with the combined results from PyI and PyII. The agreement with zero signal for the PyIII$_L$ data alone ($\chi_\nu^2$ = 327.3/96), as well as all of the combined Python data ($\chi_\nu^2$ = 516.9/127), is poor.

For PyIII$_S$, no sky position was measured by two detectors within any individual scan. However, two separate groups of scans were made at different times at each elevation offset so that each sky position was measured by a left- and right-hand detector. Within each group, an average value for every sky position was determined in the same manner as the PyIII$_L$ scans. At each elevation, the two different groups with overlapping sky measurements were then combined by forming a weighted average using the statistical errors of the individual measurements as the weighting factors. The resulting 154 measurements are shown in Figure 2, and are in poor agreement ($\chi_\nu^2 = 395/154$) with zero signal.

The entire analysis was repeated using a quadrature lockin demodulation which should have no response to a sky signal. The results for both PyIII$_L$ ($\chi_\nu^2 = 104/96$) and PyIII$_S$ ($\chi_\nu^2 = 182/154$) are significantly lower than the optical phase analysis, and are convincing evidence for a celestial source of the observed signal.

## 6. LIKELIHOOD ANALYSIS

### 6.1. BAND POWER AND GACF

A separate likelihood analysis (Readhead et al. 1989) was performed for the PyIII$_S$ and the combined PyI, PyII, and PyIII$_L$ data sets to determine the amplitude of the sky signal consistent with the data, and to assign confidence intervals to the results. A full correlation matrix, including off-diagonal terms, takes into account correlations between the measured sky positions due to the observing strategy and the assumed model (Vittorio et al. 1991).

Two measures of the anisotropy in sky temperature are presented in Table 2 for these data sets. The first model assumes that the anisotropies are described by a Gaussian autocorrelation function (GACF). For each data set, the rms amplitude, $\sqrt{C_0}$, and coherence angle, $\Theta_c$, are varied until the likelihood is maximized. The greatest sensitivity to fluctuations is at $\Theta_c = 1.0°$ for observations made with the 2.75° chop (the combined PyI, PyII, and PyIII$_L$ data set), and at $\Theta_c = 0.5°$ for the 0.92° chop data (PyIII$_S$)

The second model assumes a flat anisotropy power spectrum with $C_\ell \propto 1/\ell(\ell+1)$, and only one free parameter, the amplitude. The effective centers of the window functions are $\ell_e = 92^{+37}_{-24}$ for the combined PyI, PyII, and PyIII$_L$ data set, and $\ell_e = 177^{+66}_{-58}$ for PyIII$_S$.

## 6.2. SPECTRAL INDEX DETERMINATION

The complete Python data set, incorporating all three years of observations and with sensitivity at two angular scales, was used in a third likelihood analysis similar to Netterfield et al. (1996) to determine the amplitude and spectral index of the angular power spectrum of the measured anisotropies. The amplitude of the anisotropy, $\delta T_\ell$, is expanded as a power law in angular scale, $\ell$, as $\delta T_\ell = \delta T_{\ell_e} (\ell/\ell_e)^m$. The amplitude, $\delta T_{\ell_e}$, and spectral index, m, are then varied until the likelihood is maximized. The best-fit amplitude, $\delta T_{\ell_e} = 53^{+6}_{-4}$ μK, is specified at $\ell_e = 145$, the effective center of the window function for the combined data set. The best-fit spectral index, m = $0.02^{+.16}_{-.16}$, is consistent with a flat power spectrum between $\ell = 113$ and $\ell = 246$, the half-power points of the effective window function. The quoted uncertainties for the best-fit amplitudes and spectral index represent a 1σ confidence interval in the likelihood.

The confidence intervals quoted in Table 2 do not include the 20% systematic calibration uncertainty even though this must be included when comparing these results to other experiments and theories. Within this experiment, a calibration uncertainty only changes the overall normalization of all of the data. Because the PyIII$_L$ and PyIII$_S$ data sets have the same systematic calibration uncertainty, this uncertainty does not affect the determination of the slope m.

## 6.3. BEAM UNCERTAINTY EFFECTS

Uncertainties in the telescope pointing and chopper throw will affect the results determined in the likelihood analyses. The combined absolute and relative pointing uncertainty is estimated to be 0.15°. Slow drifts in the DC offset and amplitude of the chopper throw introduce an additional uncertainty of 0.1° in the beam positions on the sky. When convolved with our beams, these pointing uncertainties can be approximated by increasing our nominal beam widths 4 arc minutes. To estimate the effect of these uncertainties on our final results, the analysis was repeated using the wider beams.

For $PyIII_L$, the increased beam widths have no effect on the final results. However, because the beam widths and chopper throw for $PyIII_S$ are so similar, the uncertainty in the beam positions has a greater effect. A larger beam shifts to higher values the best-fit amplitude for the observed fluctuations. For $PyIII_S$ alone, the most likely band power estimate increases 10% to 64 µK. For the complete Python data set, the best-fit amplitude of the fluctuations is $\delta T_{\ell_e} = 57$ µK, an increase of 8%. The best-fit spectral index is m = 0.11. These results have been used to increase the errors listed in Table 2 to account for the effects introduced by the imperfect overlap of our beams on the sky.

## 7. FOREGROUND CONTAMINATION

Possible Galactic foreground contributions to the observed signals include emission from cool dust, as well as free-free and diffuse synchrotron emission. However, as discussed in Dragovan et al. (1994), the expected signals in our band from sources of free-free and synchrotron emission are much lower than the actual signals observed.

To estimate the contribution from dust emission to the observed signals, we simulate the Python observing strategy on the DIRBE 240 µm map of the Python fields extrapolated to 90 GHz using the COBE dust model (Wright et al. 1991) ($\epsilon_{dust} \propto \nu^n$, n = 1.65, $T_{dust}$ = 23.3 K). We find the maximum dust signal at any of the observed sky positions corresponds to an antenna temperature of less than 2 µK, and conclude that cool dust emission is not contaminating the Python data sets.

Of more concern is potential contamination by unresolved point sources. A search of the Parkes Survey catalog (Gregory et al. 1994; Wright & Otrupcek 1990) reveals 17 point sources within the 5.5° × 22° region of sky mapped by this experiment. The brightest of these has a flux of 2.7 Jy at 90 GHz, extrapolated from 22 GHz using a spectral index of β = 0.416, based on the 5 GHz and 22 GHz fluxes found in the Survey. This corresponds to an antenna temperature of 51 µK if the source was in our main beam.

To estimate the level of point source contamination, the expected amplitude of all 17 known point sources in this region was similarly estimated. A map of these sources on the sky was made, and the Python observing strategy simulated on it. Reanalyzing the data with the estimated source contributions subtracted has less than a 2% effect on the final results. We conclude that the Python data sets are not seriously contaminated by foreground point sources.

## 8. CONCLUSIONS

A $5.5° \times 22°$ region of sky has been densely sampled with greatest sensitivity to anisotropy at degree angular scales. Separate likelihood analyses of two data sets show positive detections of sky brightness variations. The total sky rms for a Gaussian autocorrelation function is $74^{+22}_{-17}$ µK at $\Theta_C = 1.0°$ and $76^{+20}_{-17}$ µK at $\Theta_C = 0.5°$. Corresponding band power results are $54^{+14}_{-12}$ µK at $92^{+37}_{-24}$ and $58^{+15}_{-13}$ µK at $177^{+66}_{-58}$.

Using a likelihood analysis of the complete set of Python observations, we find that the angular power spectrum of CMB fluctuations has a spectral index of m = $0.02^{+.18}_{-.16}$ with an amplitude of $\delta T_{\ell_e} = 53^{+13}_{-11}$ µK at $\ell_e = 145^{+59}_{-74}$ for the functional form $\delta T_\ell = \delta T_{\ell_e} (\ell/\ell_e)^m$. This result shows no evidence for differences in the level of anisotropy at angular scales between $\ell = 113$ and $\ell = 246$. In all cases, the stated uncertainties in the amplitudes and spectral index are 1 σ confidence intervals in the likelihood added in quadrature with a 20% calibration uncertainty and an estimate of the effects introduced due to imperfect overlap of the beams on the sky. The limits of $\ell$ are determined from the full width at half maximum of the window functions.

We would like to thank the Antarctic Support Associates staff for valuable assistance at the South Pole. Special thanks go to R. Pernic, D. Pernic, J. Rottman, and M. Masterman for their contributions to a successful season. G. Griffin aided in the calculation of the band power results. The COBE datasets were developed by the NASA Goddard Space Flight Center under the guidance of the COBE Science Working Group and were provided by the NSSDC. This research was supported by the James S. McDonnell Foundation, PYI grant NSF AST 90-57089, and the National Science Foundation under a cooperative agreement with the Center for Astrophysical Research in Antarctica (CARA), grant NSF OPP 89-20223. CARA is an NSF Science and Technology Center.

TABLE 1
MEAN NOISE AND OFFSETS

| Detector | PyIII$_L$ | | | PyIII$_S$ | | |
|---|---|---|---|---|---|---|
| | $\sigma_s$ (mK$\sqrt{s}$) | $\sigma_m$ (mK$\sqrt{s}$) | $\mu_{3B}$ ($\mu$K) | $\sigma_s$ (mK$\sqrt{s}$) | $\sigma_m$ (mK$\sqrt{s}$) | $\mu_{3B}$ ($\mu$K) |
| 1 | 2.3 | 3.4 | 919 | 2.2 | 2.3 | 90 |
| 2 | 2.6 | 3.3 | 880 | 2.5 | 2.5 | 104 |
| 4 | 1.4 | 2.7 | 924 | 1.4 | 1.6 | 101 |
| 5 | 6.5 | 6.1 | 483 | 6.2 | 5.6 | 168 |

NOTE.- Mean noise and offset levels. The values tabulated are averages over all scans with $\sigma_s$ and $\sigma_m$ below the cut levels discussed in the text. The 3-beam offsets ($\mu_{3B}$) are calculated by averaging across all uncut scans, irrespective of pointing position.

TABLE 2
LIKELIHOOD ANALYSIS RESULTS

| Data Set | GACF | | Band Power | | |
|---|---|---|---|---|---|
| | $\sqrt{C_0}$ | $\theta_c$ | $\sqrt{\ell_e(\ell_e+1)C_\ell/(2\pi)}$ | $\ell_e$ | $m$ |
| PyIII$_S$ | $76^{+13}_{-8}$ | $0.5°$ | $58^{+9}_{-5}$ | $177^{+66}_{-58}$ | …. |
| PyI, PyII & PyIII$_L$ | $74^{+16}_{-8}$ | $1.0°$ | $54^{+8}_{-5}$ | $92^{+37}_{-24}$ | …. |
| PyI & PyII & PyIII | …. | …. | $53^{+7}_{-4}$ | $145^{+59}_{-74}$ | $0.02^{+.18}_{-.16}$ |

NOTE.- All temperature values are in units of µK corrected for atmospheric absorption and converted from Rayleigh-Jeans to thermodynamic units. The detections are quoted at the maximum in the likelihood. The confidence intervals are determined by integrating the area under the likelihood curves and correspond to 1 sigma uncertainties in the likelihood added in quadrature with the estimated errors introduced due to uncertainties in the beam positions on the sky. The stated errors do not include the 20% calibration uncertainty. The limits of $\ell_e$ are determined from the full width at half maximum of the window functions.

FIGURE CAPTIONS

FIG. 1. - Results from the PyIII$_L$ data set are shown plotted (solid circles) with the combined results from PyI and PyII (open circles). The temperature axis is plotted in antenna temperature; multiply by 1.24 to convert to thermodynamic temperature. The six plots show the results for six rows on the sky, separated from one another by 0.92° in declination. The top row is at a declination of -51.8°. The four-beam response pattern is shown in the top panel, aligned with the point plotted at $\alpha = -20.6°$.

FIG. 2. - Results from the PyIII$_S$ data set. The same region of sky is shown as in Fig. 1. The temperature axis is plotted in antenna temperature; multiply by 1.24 to convert to thermodynamic temperature. The six plots show the results for six rows on the sky, separated from one another by 0.92° in declination. The top row is at a declination of -51.8°. The four-beam response pattern is shown in the top panel, aligned with the point plotted at $\alpha = -20.7°$.